\documentclass[twoside]{article}
\usepackage{graphicx}
\usepackage{fancyhdr}
\usepackage{multicol}
\usepackage{styl-ap}
\usepackage{natbib}

\def\nur{\nu_\mathrm{r}}

\begin{document}
\title{Distinguishing between spot and torus models of high-frequency quasiperiodic oscillations}
\author{V.~Karas\work{1}, P.~Bakala\work{2}, G.~T\"or\"ok\work{2}, M.~Dov\v{c}iak\work{1}, M.~Wildner\work{2}, D.~Wzientek\work{2},
E.~\v{S}r\'{a}mkov\'{a}\work{2}, M.~Abramowicz\work{2,3,4}, K.~Goluchov\'{a}\work{2}, G.~P.~Mazur\work{4,5}, F.~H.~Vincent\work{4,6}}
\workplace{Astronomical Institute, Bo\v{c}n\'{\i}~II~1401, CZ-14100~Prague, Czech Republic
\next
Institute of Physics, Faculty of Philosophy and Science, Silesian University in Opava, Bezru\v{c}ovo~n\'{a}m.~13,\newline CZ-74601~Opava, Czech Republic
\next
Physics Department, Gothenburg University, SE-412\,96~G\"oteborg, Sweden
\next
Copernicus Astronomical Center, ul. Bartycka 18, PL-00\,716 Warszawa, Poland
\next
Institute for Theoretical Physics, University of Warsaw, Hoza 69, PL-00\,681 Warsaw, Poland
\next
Laboratoire AstroParticule et Cosmologie, CNRS, Universit\'e Paris Diderot, 10 rue Alice Domon et Leonie Duquet, 75205, Paris Cedex~13, France}
\mainauthor{vladimir.karas@cuni.cz}
\maketitle

\begin{abstract}%
In the context of high-frequency quasi-periodic oscillation (HF QPOs) we further explore the appearance of an observable signal generated by hot spots moving along quasi-elliptic trajectories close to the innermost stable circular orbit in the Schwarzschild spacetime. The aim of our investigation is to reveal whether observable characteristics of the Fourier power-spectral density can help us to distinguish between the two competing models, namely, the idea of bright spots orbiting on the surface of an accretion torus versus the scenario of intrinsic oscillations of the torus itself. We take the capabilities of the present observatories (represented by the Rossi X-ray Timing Explorer, RXTE) into account, and we also consider the proposed future instruments (represented here by the Large Observatory for X-ray Timing, LOFT). 
\end{abstract}


\begin{multicols}{2}

\section{Introduction}
\renewcommand*{\thefootnote}{\fnsymbol{footnote}}
\footnotetext{Talk presented at 10$^{\rm th}$ INTEGRAL/BART Workshop in Karlovy Vary (Czech Republic, 22-25 April 2013).}
Low-mass X-ray binaries (LMXBs) represent a particular example of astronomical sources in which puzzling quasi-periodic modulation of the observed signal can develop and reach very high (kilohertz) frequencies \cite{Kli:2006:CompStelX-Ray:}. Orbiting inhomogeneities, a.k.a.\ spots residing on the surface of an accretion disk, have been introduced as a tentative explanation for persistant high-frequency oscillations (HF QPOs) observed in X-rays from accreting black holes and neutron stars in LMXBs. These QPOs have been discussed, in particular, in a series of papers \citep{kar:1999,mor-ste:1999,ste-vie:1998a,ste-vie:1998b,ste-vie:1999,ste-vie:2002} that explain QPOs as a direct manifestation of modes of relativistic epicyclic motion of blobs at various radii $r$ in the inner parts of the accretion disc. Within the model, the two observed HF QPO peaks arise due to the Keplerian and periastron precession of the relativistic orbits. 

In a similar manner, the authors of refs.\ \cite{cad-etal:2008,ger-etal:2009,kos-etal:2009} introduced a concept according to which the QPOs are generated by a tidal disruption of large accreting inhomogeneities. Recently, \cite{ger:2012} argued that the behaviour of  the azimuth phase $\phi(t)$ for non-closed quasi-elliptic orbits in the curved spacetime can be responsible for the observed pairs of HF QPOs.

In this contribution we investigate the behaviour of the observable signal produced by radiating small circular hotspots. We discuss the detectability of the produced signal propagated from the strongly curved spacetime region, and in particular we concentrate of the properties of Fourier power-spectral density (PDS), as predicted in different scenarios. It has been known that PDS reflects the physical properties of the source, however, it does not provide a complete description. We thus explore a possibility to discriminate via PDS between the two specific models that are thought to provide a promising scheme for HF QPOs in black hole sources. In fact, according to the Wiener-Khinchin theorem, the power spectral function $S(\omega)$ can be calculated from the autocovariance function $C(t)$ as \cite{bendat:2000}
\begin{equation}
S(\omega)={\cal F}[C](\omega).
\end{equation}
It can be seen that only the Gaussian processes are completely determined by their power spectra. Nonetheless, it is interesting to investigate the predicted profiles of the power spectrum for well-defined processes and to reveal specific features, their differences and similarities.

In our discussion we consider the capabilities of the present X-ray observatories represented by Rossi X-ray Timing Explorer (RXTE), as well as the proposed future instruments represented by the Large Observatory for X-ray Timing (LOFT). We also compare the signal produced by spots to the signal obtained from another specific kind of simulations assuming axisymmetric epicyclic disc-oscillation modes. 
Our present paper is based on a recent work \citep{bakala:2013} where further details can be found. In particular, in the cited paper we explore the width of the QPO peaks with 3:2 frequency ratio that has been reported to occur in a number of sources.

\section{Set-up of the model}
Qualtitative mathematical properties of Fourier power spectra, as predicted by the hot-spot scenario, have been studied in our refs.\ \cite{pec:2013,pec:2008}. Next,
in ref.\ \cite{kar:1999} a quantitative study of the expected PDS was presented and the orbital motion of the spots and their distribution within a narrow range of radii was discussed. The adopted phenomenological description of the source employs spots or clumps orbiting in the Schwarzschild metric as an approximation to more realistic models of inhomogeneous disc--type accretion flows around black holes and neutron stars. In order to reproduce the quality factor of the observed QPOs, it was found that spots must be distributed in a zone of only several gravitational radii from the central black hole, with their observed luminosity influenced by Doppler and lensing effects. Furthermore, we found that the expected values of the quality factor of the oscillations can reach about several hundred (typically, $Q\approx3\times10^2$). However, the assumption about strictly circular orbits was clearly a simplification which calls for further discussion in the present paper. Furthermore, the effect of background noise has to be taken properly into account
\cite{Wit:2012}.

\begin{figure*}  
%--------------------------------------------------------------------------------
\begin{minipage}{1\hsize}
\includegraphics[width=\hsize]{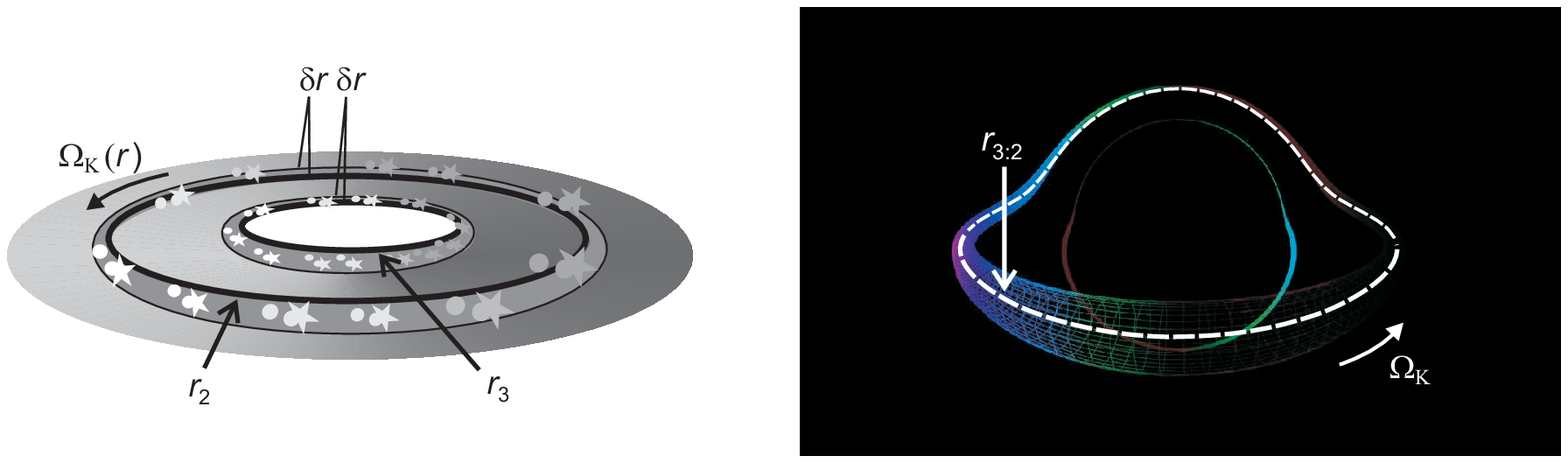}
\end{minipage}
\caption{Left: Sketch of the QPO model based on spots orbiting close to two preferred radii, producing the 3:2 ratio of observed frequencies. Right: Illustration of the role of lensing and Doppler effects in the visual appearance of torus located at the resonant orbit $r_{3:2}$. Figure adopted from ref.\ \cite{bakala:2013}.}
\label{figure:3to2models}
%--------------------------------------------------------------------------------
\end{figure*}

\begin{figure*}  
%--------------------------------------------------------------------------------
\begin{minipage}{1\hsize}
\includegraphics[width=\hsize]{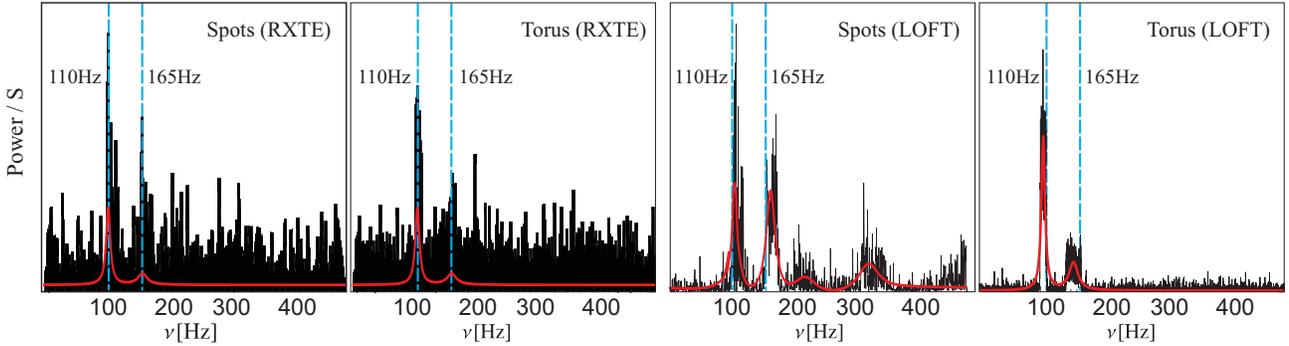}
\end{minipage}
\caption{Comparison between multiple spot and oscillating torus PDS obtained for the two instruments. Superimposed red curves indicate various multi-Lorentzian models.}
\label{figure:PDSfit}
%--------------------------------------------------------------------------------
\end{figure*}

The left panel of Figure~\ref{figure:3to2models} illustrates the spot scenario which we investigate. The radial distribution or drifting of the spots can clearly result in various levels of signal coherence. Nevertheless, small circular spots related to a single preferred radius cannot reproduce the often observed 3:2 frequency ratio. We also consider a more elaborate scheme where the multiple spots are created and drifted around radii close to two preferred orbits with Keplerian frequencies roughly in the 3:2 ratio. These orbits are set as $r_{2}=8M$ (radius where the radial epicyclic frequency reaches the maximum value) and $r_{3}=6M$ (ISCO). Spots are then created within the regions $[r_{i}-\delta{r},\,r_{i}]$ with the size given by $\delta{r}=0.75M$.  Since we assume the black hole mass $M\doteq11M_\odot$,  our setup leads to the main observable frequencies around 110Hz and 160Hz (see Figure~\ref{figure:PDSfit} drawn for the signal fraction $n=10\%$ and $\mathcal{D}=65^{\circ}$). In the following consideration we compare the PDS obtained for this setup to the PDS resulting from the model of the oscillating optically thin torus slowly drifting through the resonant radius $r_{3:2}$. 

To define the torus kinematics, we assume the $m=0$ radial and vertical oscillations with equal intrinsic amplitudes. The possible QPO origin in the resonances between this or similar disc oscillation modes has been extensively discussed in works of \cite{abr-klu:2001,abr-etal:2003a,abr-etal:2003b,hor:2005,klu-etal:2004,tor-etal:2005}, and several other authors. Here we adopt the concept previously investigated by \cite{bur-etal:2004} who focused on optically thin torus with slender geometry. 

The visual appearance of torus influenced by lensing and Doppler effects is illustrated in the right panel of Figure~\ref{figure:3to2models}.  Within the adopted concept the periodic changes of the observed luminosity are partially governed by the radial oscillations due to changes of the torus volume while the vertical oscillations modulate the flux just due to lensing effects in the strong gravitational field. 

The contribution of the two individual oscillations to the variations of the observed flux thus strongly depends on the inclination angle \citep[see also ref.][]{maz-etal:2013}. Here we set $\mathcal{D}=65^{\circ}$ where the fractions of the power in the two observed peaks are comparable. We set the black hole mass $M=5.65M_{\odot}$ and $a=0$ $(r_{3:2}=10.8M)$, implying that the two oscillatory frequencies are $\nu_\theta(r_{3:2})=160$Hz and $\nu_\mathrm{r}(r_{3:2})\doteq110$Hz. Assuming this setup we produce torus drift lightcurves for the interval $r/r_{3:2}\in[0.97,1.03]$. The resulting PDS drawn for the signal fraction $n=10\%$ is included  in Figure~\ref{figure:PDSfit}. We note that similar PDS can be reached assuming e.g. a near extreme rotating black hole with $a=0.98$ and $M\doteq18M_{\odot}$.  

Using Figure~\ref{figure:PDSfit} we can finally confront the predictions obtained for spots drifted around preferred radii to those expected for the oscillating torus slowly passing the resonant orbit $r_{3:2}$. Inspecting this Figure, we can find that the RXTE PDS obtained for the given setup of the two models are rather similar. On the other hand, the LOFT PDS clearly reveal presence/absence of the harmonics additional to the 3:2 peaks representing the signature of spot motion.

%--------------------- FIGURE ----------------------------------------------------
%--------------------------------------------------------------------------------
%--------------------------------------------------------------------------------

%--------------------- FIGURE ----------------------------------------------------
%--------------------------------------------------------------------------------

We assume the global source flux described by the approximations of the spectral distribution $N(E)$ and power density spectra $P(\nu)$,
%-------------------------
\begin{eqnarray}
\label{equation:global:spectrum}
&&N(E)= kE^{-2.5},\\
%-------------------------
\label{equation:global:variability}
&&P(\nu)= p_0\nu^{p_1}+ \frac{1}{\pi}\frac{p_3p_4}{(\nu-p_2)^2+p_3^2},
%-------------------------
\end{eqnarray}
%-------------------------
where $k$ is chosen to normalize the assumed countrate roughly to $1$Crab and $p_{\mathrm{i}}=[0.001,-1.3,2.5,0.8,0.002]$. This setup roughly corresponds to the so-called high steep power law (HSPL) state in GRS~1915+105. 

%---------------------------------------------------------
\section{Expected properties of PDS}
%---------------------------------------------------------

%--------------------- FIGURE ----------------------------------------------------
\begin{figure*}  
%--------------------------------------------------------------------------------
\begin{minipage}{1\hsize}
\includegraphics[width=\hsize]{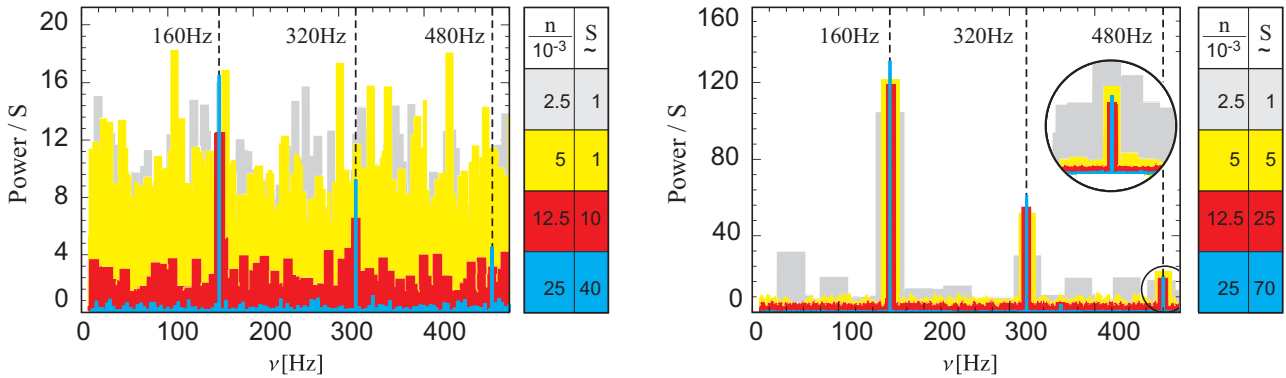}
\end{minipage}
\caption{The resulting PDS obtained for different levels of the signal fraction $n$ assuming $D=80^{\circ}$ and $\mathcal{E}=1M$. For the sake of clarity, the individual PDS are rescaled by unifying factor S. \small {{Left:}} Outputs considering the RXTE capabilities. {{Right:}} Outputs considering LOFT capabilities. One should note that the lowest displayed values of $n$ corresponding to gray and yellow colour do not indicate any significant features within the RXTE PDS. On the other hand, the LOFT PDS already reveal Keplerian frequency (gray and yellow PDS) respectively its first two harmonics (yellow PDS).}
\label{figure:PDSD80}
%--------------------------------------------------------------------------------
\end{figure*}
%--------------------------------------------------------------------------------

%--------------------- FIGURE ----------------------------------------------------
\begin{figure*}
%--------------------------------------------------------------------------------
\begin{minipage}{1\hsize}
\begin{center}
\begin{minipage}{1\hsize}
\includegraphics[width=\hsize]{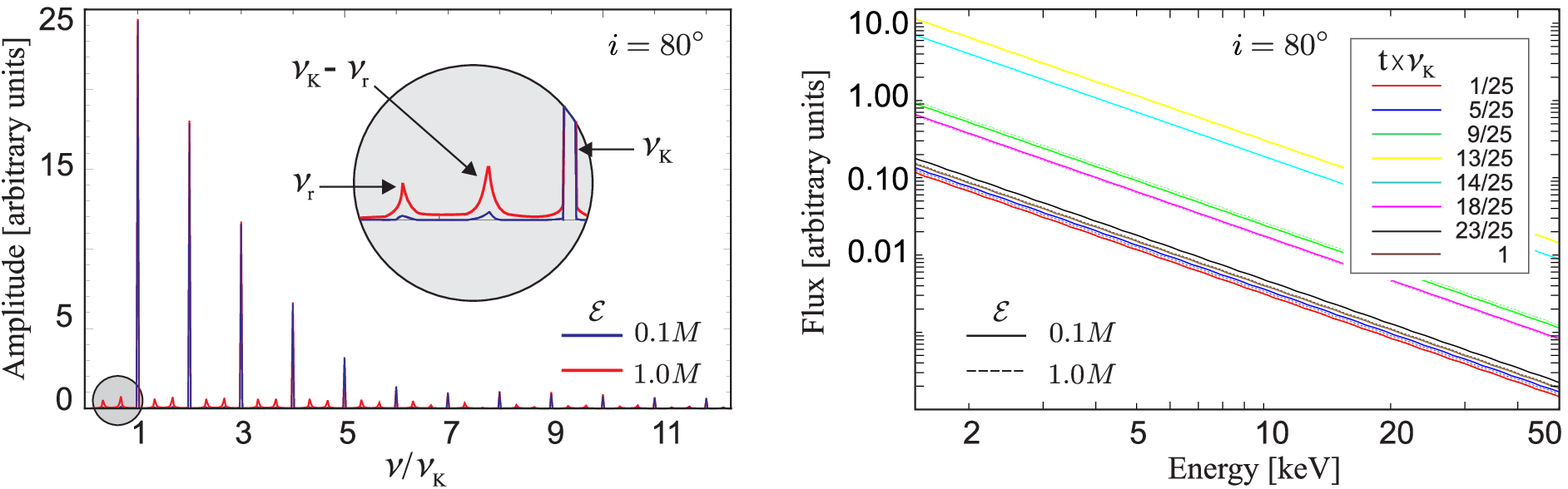}
\medskip

\noindent
\includegraphics[width=\hsize]{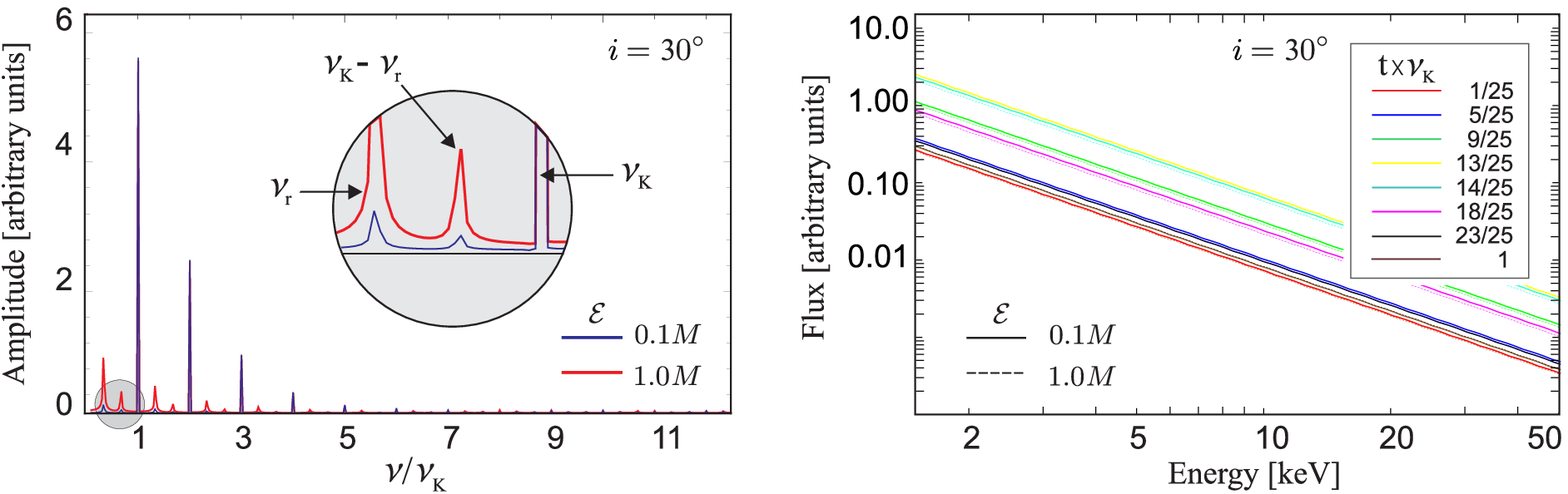}
\end{minipage}
\end{center}
\end{minipage}
\caption{Expected net spot flux measured by a distant observer for different inclination angles. {Left:} Amplitude spectrum. {{Right:}} Time dependent energy spectra drawn for the distant observer.}
\label{figure:D30}
%--------------------------------------------------------------------------------
\end{figure*}
%--------------------------------------------------------------------------------
%--------------------- FIGURE ----------------------------------------------------
\begin{figure*}  
%--------------------------------------------------------------------------------
\begin{minipage}{1\hsize}
\includegraphics[width=\hsize]{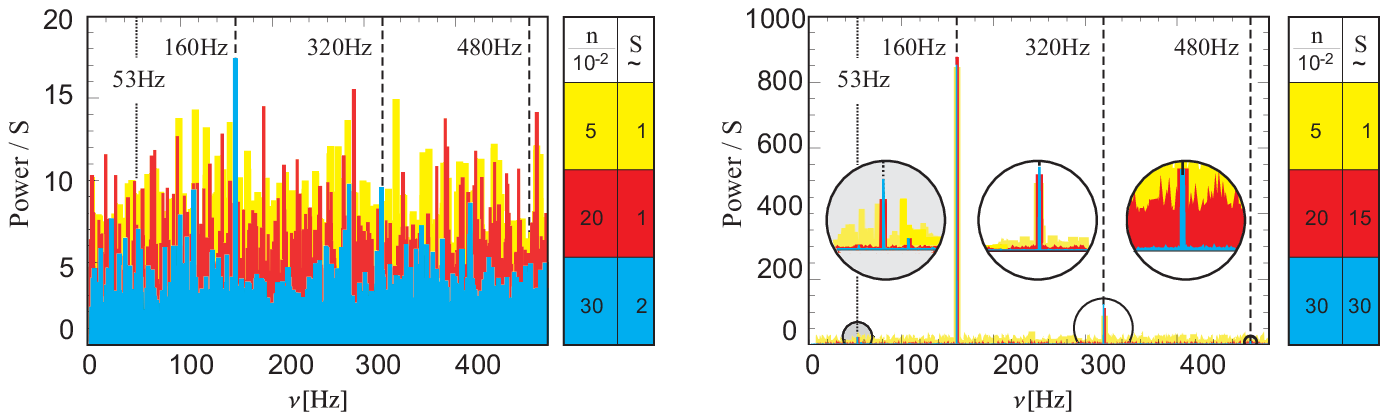}
\end{minipage}
\caption{The resulting PDS obtained for different levels of $n$ assuming $D=30^{\circ}$ and $\mathcal{E}=1M$. For the sake of clarity, the individual PDS are rescaled by unifying factor S. \small {{Left:}} Outputs considering RXTE capabilities. {{Right:}} Outputs considering LOFT capabilities. We note that RXTE PDS includes a barely significant excess of power at 160Hz only for the highest displayed value of $n$. For the same value, the LOFT PDS reveal the first two harmonics and also the radial epicyclic frequency (53Hz).}
\label{figure:PDSD30}
%--------------------------------------------------------------------------------
\end{figure*}
%--------------------------------------------------------------------------------

Figure~\ref{figure:PDSD80} shows PDS resulting from the RXTE and LOFT simulations assuming various levels of $n$ and the inclination $\mathcal{D}=80^{\circ}$ (i.e., nearly equatorial view). Since the signal from the spot strongly depends on the source inclination $D$ \citep[e.g.\ ref.][]{sch-ber:2004a}, we will compare the results for two representative values of $D$ corresponding to the nearly equatorial view $D=80^{\circ}$ and the view close to vertical axis $D=30^{\circ}$. 

In the following, we assume a spot orbiting at $r=6.75M$ with constant angular velocity of the Keplerian value $\Omega_\mathrm{K}$. The spot trajectory slightly deviates from the circular shape due to the radial epicyclic oscillation, having small amplitude $\mathcal{E}>0$.

\subsection{Nearly equatorial (edge-on) view} 
In Figure~\ref{figure:D30} (upper panels) we include amplitude spectra and time dependent energy spectra of the net spot signal calculated for the distant observer at $D=80^{\circ}$. The spot signal is dominated by the Keplerian frequency and its harmonics amplified by relativistic effects which is well illustrated by the amplitude spectrum on the upper left panel of Figure \ref{figure:D30}. The eccentricity corresponding to the amplitude of the radial epicyclic oscillation $\mathcal{E}=0.1M$ causes only negligible modulation at the radial and precession frequencies. The increased eccentricity corresponding to the amplitude of the radial epicyclic oscillation $\mathcal{E}=1M$  can be well recognized in the amplitude spectra but the signal is still dominated by the Keplerian frequency and its harmonics. The time dependent energy spectra of the spot are depicted in the upper right panel of Figure~\ref{figure:D30}. We can see that they clearly reveal the signatures of relativistic redshift effects.

So far we have reproduced just the variability and spectra of the net spot flux. In order to assess the observable effects we have to study the total composition of the net spot flux together with the global source flux given by Equations (\ref{equation:global:spectrum}) and (\ref{equation:global:variability}). Assuming this composed radiation, we can consider the capabilities of the RXTE and LOFT instruments using their response matrices and provided software tools. The time-dependent spectra describing the composed radiation are then convolved with the appropriate response matrix giving an estimate of the observed data. These are Fourier transformed to the resulting power spectra. Within such consideration the detectability of the spot signatures depends obviously on the fraction of photons from the spot in the total flux. We referred to this {\em signal to noise ratio} shortly as the signal fraction $n$.

Figure~\ref{figure:PDSD80} includes PDS resulting from the RXTE and LOFT simulations assuming various levels of $n$ and the inclination $\mathcal{D}=80^{\circ}$. It includes the cases when the signal is weak for the RXTE and there are no significant features within its PDS as well as the high signal fraction when first two harmonics of the Keplerian frequency can be seen. Comparing the both panels of this Figure we can deduce that when the weak QPO signal corresponding to the hot-spot Keplerian frequency is around the limits of the RXTE detectability, the LOFT observations can clearly reveal its first and second harmonics. We checked that there is, in practice, no qualitative difference between the cases of $\mathcal{E}=0.1M$ and $\mathcal{E}=1M$. It is therefore unlikely that the periastron precession or radial epicyclic frequency can be detected in addition to the harmonics when the inclination angle is close to the equatorial plane.

\subsection{View close to vertical axis} 
For $D=30^{\circ}$ (lower panels), the signal is dominated by the Keplerian frequency but the harmonics are much less amplified in comparison to the nearly equatorial view (see the bottom left panel of Figure~\ref{figure:D30}). Eccentricity corresponding to the amplitude of radial epicyclic oscillation $\mathcal{E}=0.1M$ causes again rather negligible modulation at the radial and precession frequencies. Nevertheless, we can see  that the increased eccentricity of $\mathcal{E}=1M$ affects the variability more than for the large inclination angle. Furthermore, its influence is comparable to those of second harmonics of the Keplerian frequency. 
The time dependent energy spectra are again depicted in the lower right panel of Figure~\ref{figure:D30}.

Finally, figure~\ref{figure:PDSD30} shows PDS resulting from RXTE and LOFT simulations assuming various levels of $n$. It is drawn for $\mathcal{E}=1M$ and includes the few cases when the signal is weak for the RXTE and there are no significant features within its PDS, plus one case when some feature at the Keplerian frequency can be seen. Comparing the both panels of this Figure we can deduce that when the weak QPO signal corresponding to the hot-spot Keplerian frequency is around the limits of the RXTE detectability, the LOFT observations can clearly reveal its first and second harmonics but also the radial epicyclic frequency $\nur$. Although it is not directly shown, we check that decreasing of eccentricity to $\mathcal{E}=0.1M$ leads to similar PDS but with the missing feature at $\nur$.

\section{Conclusions}
We can identify the signatures of the spot motion mostly with the harmonic content of the observable signal. For large inclination angles, the LOFT observations could easily reveal the Keplerian frequency of the spot together with its first and second harmonics when the strongest (but weak) single signal is around the limits of the RXTE detectability. Nevertheless, radial epicyclic frequency could be also found providing that the inclination is small. In our analysis we have paid attention to the timing signatures of the motion of small circular spots radiating isotropically from the slightly eccentric geodesic orbits. The case of highly eccentric orbits and/or spot having large azimuthal shear will be presented elsewhere.

We studied the comparison between spot and torus scenarios serving as  fiducial representations of two very specific kinematic models. Obviously, any general validity of this discussion is limited. For instance, a consideration of resonance driven effects or the role of torus geometrical thickness could give rise to some harmonic content in the signal from the oscillating tori. Despite uncertainties, the elaborated comparison indicates clearly that the increased sensitivity of the proposed LOFT mission can be crucial for resolving the QPO nature. We refer the reader to ref.\ \cite{bakala:2013} for additional details.

%\begin{myfigure}
%\centerline{\resizebox{70mm}{!}{\includegraphics{figure.eps}}}
%\caption{Name of figure 1}
%\label{author-fig1}
%\end{myfigure}

%\begin{figure}[htb]
%\centerline{\includegraphics{figure.eps}}
%\caption{Big figure}
%\label{author-fig2}
%\end{figure}

\thanks
We acknowledge the project CZ.1.07/2.3.00/20.0071 Synergy supporting the international collaboration of Institute of Physics in Opava,
the Czech Science Foundation project GA\v{C}R 13-00070J in Prague (VK), GA\v{C}R~209/12/P740 (GT, E\v{S}), 
and the European Union 7th Framework Programme No. 312789 `StrongGravity' (MD). 
We also acknowledge the Polish grant NCN UMO-2011/01/B/ST9/05439 and the Swedish VR grant (MAA).
Astronomical Institute is supported by the research program RVO:67985815.

\end{multicols}
\end{document}